\documentclass[preprint,12pt]{elsarticle}
\usepackage{amssymb}
\usepackage{amsmath}
\usepackage{booktabs}
\usepackage{siunitx}
\usepackage{subcaption}
\usepackage{graphicx}
\usepackage{dcolumn}
\usepackage{bm}
\usepackage{hyperref}

\newcommand{\BV}{Brunt-V\"{a}is\"{a}l\"{a}}

\journal{Int. J. Multiphase Flow}

\begin{document}

\begin{frontmatter}

\title{Hydrodynamic coupling and retention time of inertial spheres crossing and bouncing at density interfaces}

\author[tau]{Chen Mortenfeld}
\author[icl]{Maarten van Reeuwijk}
\author[tau]{Aviv Littman}
\author[tau]{Alex Liberzon\corref{cor1}}
\ead{alexlib@tauex.tau.ac.il}

\cortext[cor1]{Alex Liberzon}
\address[tau]{Turbulence Structure Laboratory, School of Mechanical Engineering, Tel Aviv University, Israel}
\address[icl]{Department of Civil and Environmental Engineering, Imperial College London, UK}

\begin{abstract}
Inertial spheres settling through sharp density interfaces can arrest, reverse direction, and resume descent, a phenomenon known as bouncing. Using synchronized particle image velocimetry and tracking in water-salt and water-glycerol stratifications, we demonstrate that bouncing is the dynamic response of a coupled sphere-fluid composite. As the sphere crosses the interface, it entrains a boundary layer of lighter fluid, creating a transient buoyant wake.

We formalize this mechanism into a phenomenological dynamic model that couples the momentum of the sphere with the entrainment and detachment of the wake. Evaluating the stationary points of this system yields a criterion that classifies trajectory archetypes (smooth crossing, deep minima, and bouncing) across different fluid regimes. We identify a dual role of viscosity, which is often overlooked by density-only models: it acts kinematically to thicken the boundary layer and increase the entrained wake volume, and dynamically to alter the drag-to-weight balance.

Furthermore, we describe the spatial dynamics of the crossing: inertia-dominated spheres penetrate further into the lower fluid before arresting due to a longer wake-detachment length, whereas buoyancy-dominated spheres arrest closer to the interface. Finally, we show that the retention time is governed by the buoyancy-driven detachment of the entrained film. By normalizing the measured retention times with a characteristic detachment timescale, we collapse the data from different viscosity regimes onto a single curve. These physical insights allow the prediction of trajectory archetype, deceleration depth, and retention time from bulk properties.
\end{abstract}




\end{frontmatter}

\section{Introduction}
\label{Section : Introduction}

In a wide range of environmental and industrial fluid mechanics problems, density stratification arising from temperature, salinity, or composition gradients profoundly influences the transport of submerged particles~\cite{Abaid2004}, droplets~\cite{Li2019}, and bubbles~\cite{Mandel2020, Blanchette2012}. Retention time, the duration a particle spends moving across a density interface, is an important operational parameter for biological, chemical, and physical processes~\cite{Mandel2020}: it governs the formation of marine snow, the trapping of microplastics at haloclines, and the efficiency of industrial settling tanks. In our experiments, retention time varies from a couple of seconds to minutes with only small changes in sphere or fluid properties, signaling a sharp, fluid mechanics mechanism-driven transition that existing models cannot fully predict.

\begin{figure}[!ht]
	\centering
	\includegraphics[width=.85\textwidth]{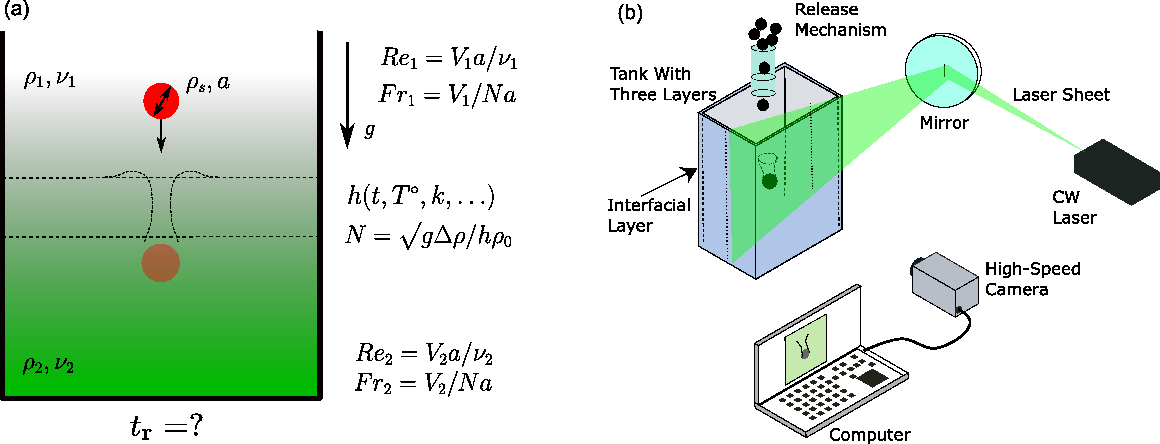}	
	\caption{(a) Schematic formulation of the physical setup. Notations appear in the text. (b) Schematics of the experimental setup.}
	\label{Figure: Experimental Setup}
\end{figure}

The phenomenon now called bouncing was first reported and modeled by Abaid~\cite{Abaid2004}, who showed that inertial spheres settling through a sharp two-layer interface can arrest, reverse direction, and ultimately resume descent. The effect depends on sphere and fluid properties in ways that resisted simple prediction: spheres denser than both layers still reversed, while nominally similar spheres in neighboring parameter sets did not. Similar temporary motion reversal has been observed for liquid drops~\cite{Li2019} and deformable bubbles at free surfaces \cite{Gao2025}. While capillary effects and Marangoni stresses drive the rebound in those systems, they share a common conclusion with rigid spheres: fluid properties beyond static density, particularly viscosity, fundamentally govern the interface crossing dynamics. While previous experiments~\cite{Srdic-Mitrovic1999} did not observe this phenomenon in their apparently similar parameter range, subsequent studies~\cite{Verso2019, Zhang2019, Wang2024} confirmed its existence across a range of sphere sizes and density jumps, and established the basic flow picture: the sphere drags a wake of lighter fluid across the interface, that wake detaches and returns as a buoyant plume, and the interface deforms in response. The problem at hand is schematically depicted in \autoref{Figure: Experimental Setup}a. Here, 1 and 2 denote the two fluid layers, with layer 1 being the lighter upper fluid in which the sphere begins its descent, or a lighter particle or oil droplet begins its ascent~\cite{Verso2019, Mandel2020}. In this sense, the notation 1,2 is more general than the notation of upper/lower as adopted by \cite{Wang2024}, as it implies that the same mechanisms act on both positively and negatively buoyant spheres. The dimensionless parameters of the problem include the Reynolds numbers $Re_{1,2} = U_{1,2}\,a/\nu_{1,2}$, the Froude numbers $Fr_{1,2} = U_{1,2}/(Na)$, and the interface-to-sphere diameter ratio $a/h$. Here $N = \sqrt{g\Delta\rho/(h \rho_0)}$ is the \BV~frequency and $\rho_0 = (\rho_1+\rho_2)/2$.

Three modeling frameworks have attempted to predict which spheres bounce. Abaid et al.~\cite{Abaid2004} provided the first model: a virtual added mass plus an empirical spring constant and some artificial spring length that reproduces trajectories but has no clear physical derivation, so the restoring force has no counterpart in the Navier-Stokes equations for a viscous fluid. Their system of four coupled ODEs requires at least five empirical parameters fitted independently for each trajectory, making it descriptive rather than predictive. Camassa et al.~\cite{Camassa2022} used the equations of motion with a quasi-static potential energy assumption, deriving the critical density triplet condition for arrest at the interface:
\begin{equation}\label{eq:critical_density}
\frac{\rho_s}{\rho_1} \leq a_1 \frac{\rho_2}{\rho_1} + a_2,
\end{equation}
where $a_1$, $a_2$ are empirical coefficients. Wang et al.~\cite{Wang2024} extended the experiments to larger spheres, added numerical simulations, and proposed a Reynolds-based empirical value that separates bouncing spheres from non-bouncing ones. The three frameworks work within their calibration range, yet they largely rely on static density thresholds or empirical Froude/Reynolds scalings~\cite{Camassa2022, Wang2024}. Crucially, they fail to predict behavior when a significant viscosity jump accompanies the density interface, as observed in our water-glycerol experiments.

We hypothesize that viscosity is not merely a modifier to terminal velocity, but plays a dual, structural role in the physics of interface crossing. First, it acts \emph{kinematically}: a higher viscosity in the upper layer thickens the sphere's boundary layer, allowing it to entrain a larger volume of lighter fluid, effectively dragging a buoyant wake into the lower layer. Second, it acts \emph{dynamically}: the lower layer's viscosity dictates the drag force that resists the sphere's inertia.

Furthermore, existing models do not explain the spatial dynamics of the crossing, specifically, why some spheres arrest exactly at the density interface, while denser spheres plunge many diameters deep into the lower fluid before reaching their minimum velocity. By treating the sphere and its entrained wake as a co-moving Lagrangian composite body, we aim to unify these observations, explaining not only \emph{if} a sphere will bounce, but \emph{where} it will arrest, and \emph{how long} it will be retained.

To close these gaps and demonstrate that the entrained light fluid model of \cite{Verso2019} can be extended to the stopping and bouncing regime, we deliberately designed a new set of experiments with larger spheres, larger density contrasts, moderate $Re_1$, two fluid systems, and synchronized fluid/sphere visualization through PIV/PTV. This parameter range is close to the work of \cite{Wang2024} and allows us to relate our findings to the theoretical framework of \cite{Camassa2022a} and the mechanistic model of \cite{Abaid2004}.

Section~\ref{Section : Materials} describes the experimental setup and sphere and fluid properties. Section~\ref{subsection:first_principles} develops a coupled ODE system for sphere velocity $U$ and entrained film volume $V_e$. Section~\ref{Section : Results} presents the experimental results, first in the raw format, and then scaled according to the previous theoretical frameworks, emphasizing the viscous effects and discrepancies, followed by an applicative recipe that predicts trajectory type and retention time from bulk fluid and sphere properties alone. Conclusions and open questions are in Section~\ref{Section: Conclusions}.

\section{Materials and methods}
\label{Section : Materials}

We conducted experiments focusing on the mechanisms that control particle behavior in stratified fluids, specifically those that exhibit significant slowdown, arrest and bouncing. We utilize a two-fluid system with viscosity and density jumps, observing bouncing in a previously unexplored regime.

The interface thickness, $h$, increases due to molecular diffusion on the order of hours. Molecular diffusion $k$ strongly depends on the fluid temperature, $T$, which must be controlled carefully. Because separate experimental runs were performed with short series of spheres in a temperature-controlled environment, we assume the sphere diameter to interface ratio, $a/h$, measured for each run at the beginning of the run, is constant within each experimental run. We work with spheres practically of the same diameter, so this ratio varies only between different experimental days. We list all the experimental parameters in Table~\ref{Table : Combined Parameters}. 

\subsection{Experimental Design and Parameter Space}
\label{Section : Setup}

The experiment setup is shown schematically in \autoref{Figure: Experimental Setup}. The experimental apparatus consists of a glass tank ($0.15 \times 0.3$ m cross-section, 0.6 m depth) with two protected sides to minimize laser reflections and improve imaging contrast. The key factor in our setup is the simultaneous use of sphere and tracer image velocimetry (PIV/PTV), which enables direct observation of sphere and fluid motion.

To explore the role of fluid properties systematically, we conducted two distinct series of experiments:
\begin{itemize}
    \item Series \textbf{A}: Water-salt stratification
    \item Series \textbf{B}: Water-glycerol stratification
\end{itemize}

In series A, the density ratios $\rho_{s}/\rho_{1,2}$ and $\Delta \rho = \rho_{2} - \rho_{1}$ closely matched those reported in previous studies~\cite{Abaid2004, Camassa2022, Wang2024}. In series B, we systematically increased the density ratios and varied the fluid viscosities, $\nu_{1,2}$. The viscosity settings yield new insights into the interaction mechanisms of settling spheres across density stratified layers.

As shown in Table~\ref{Table : Combined Parameters}, our experiments span Reynolds numbers $Re_1 = 106.6-438.8$ (upper layer) and $Re_2 = 1.1-264.3$ (lower layer), in some cases extending the regimes of previous studies. We achieved the extended range using different combinations of stratified fluid layers with higher viscosity and lower Reynolds number, observing bouncing in an almost order of magnitude larger density ratios, $(\rho_s - \rho_1)/\rho_2 \approx 0.05$, compared to previous works of \cite{Abaid2004, Camassa2022, Wang2024}. Several of our experimental runs for the lower end of density ratio $\Delta \rho$ are close to the parameter range of \cite{Wang2024}, allowing us to compare our results directly.  

\begin{table}[!ht]
    \centering
    \begin{tabular}{c c c c}
    \toprule
    Type & Diameter (mm) & Density (kg m$^{-3}$) & Material \\
    \midrule
    P$_1$ & 9.546 $\pm$ 0.021 & 1133 $\pm$ 10 & Nylon \\
    P$_2$ & 10.024 $\pm$ 0.012 & 1109 $\pm$ 12 & Nylon \\
    P$_3$ & 9.769 $\pm$ 0.095 & 1214 $\pm$ 22 & Cellulose Acetate \\
    \bottomrule
    \end{tabular}
    \caption{Sphere characteristics with 95\% confidence intervals. Manufacturers: Salem Specialty Ball Co. and Cospheric Inc., Santa Barbara.}
    \label{Table : sphere attribute}
\end{table}

\begin{table}
   \centering
   \caption{Parameter ranges for Series A and B experiments}
   \label{Table : Combined Parameters}
   \begin{tabular}{l | c c c c}
   \toprule
   Parameter & Symbol & Series A & Series B & Units \\
   \hline    
   \midrule
   Upper layer density & $\rho_1$ &  1090--1120 & 1081--1133 & kg m$^{-3}$ \\
   Lower layer density & $\rho_2$ & 1100--1130 & 1096--1164 & kg m$^{-3}$ \\
   Upper layer viscosity & $\nu_1$ &  1.317--1.848 & 2.689--6.703 & 10$^{-6}$ m$^2$ s$^{-1}$ \\
   Lower layer viscosity & $\nu_2$ &  1.415--1.937 & 3.477--12.79 & 10$^{-6}$ m$^2$ s$^{-1}$ \\
   Interface width & $h$ &  0.01--0.05 & 0.01--0.02 & m \\
   Upper Reynolds number & $Re_1$  & 174.0--438.8 & 106.6--331.7 & - \\
   Lower Reynolds number & $Re_2$  & 1.1--264.3 & 16.0--150.3 & - \\
   Upper Froude number & $Fr_1$ & 1.6--5.1 & 1.4--4.1 & - \\
   Lower Froude number & $Fr_2$ & 0.0--3.4 & 0.2--2.6 & - \\
   \BV~freq & $N$ & 1.2--2.7 & 2.3--4.0 & s$^{-1}$ \\
   \bottomrule
   \end{tabular}
\end{table}

\subsection{Fluid Preparation and Stratification}

We established the stratified layers using a carefully controlled bottom-filling technique \cite{Verso2019, Srdic-Mitrovic1999, Wang2024}. We first fill the top layer with the lighter solution. After a specific time, when the flow stops completely, the denser solution is introduced through a bottom port at approximately 500 ml/min, regulated by a pump with a control valve. This approach minimizes mixing while establishing the density interface. The transition layer between fluids develops naturally through molecular diffusion.

All experiments were conducted in an enclosed, continuously temperature controlled room to avoid temperature variations that could potentially alter the fluid density, viscosity, and molecular diffusivity~\cite{Wang2024}. Fluid properties were measured before each experiment using calibrated instruments: density, $\rho$: Mettler Toledo Easy D40 density meter (accuracy $\pm$0.5 kg m$^{-3}$), viscosity $\mu$: Ostwald Plattern calibrated viscometer (uncertainty 3\% for Series A, 5\% for Series B), and the temperature was monitored continuously during the preparation and experiments using a laboratory-grade liquid thermometer.

\subsection{Interface Characterization}

We employed two complementary methods for characterizing the density interface. For the series A experiments, the interface position was determined from careful density measurements and volume tracking during filling and was verified in control experiments. For series B, we measured vertical density profiles directly using fluid samples at precise depths, then analyzed them with the density meter.

Because both fluid pairs (A: water--salt and B: water--glycerol) are fully miscible, there is no surface tension at the density interface and capillary effects are absent. The rigid spheres are released from a submerged depth in the lighter fluid and do not exhibit interfacial tension with either fluid, so the dynamics are governed entirely by buoyancy, drag, and entrainment.

\begin{figure}[!ht]
    \centering
    \includegraphics[width=0.95\textwidth]{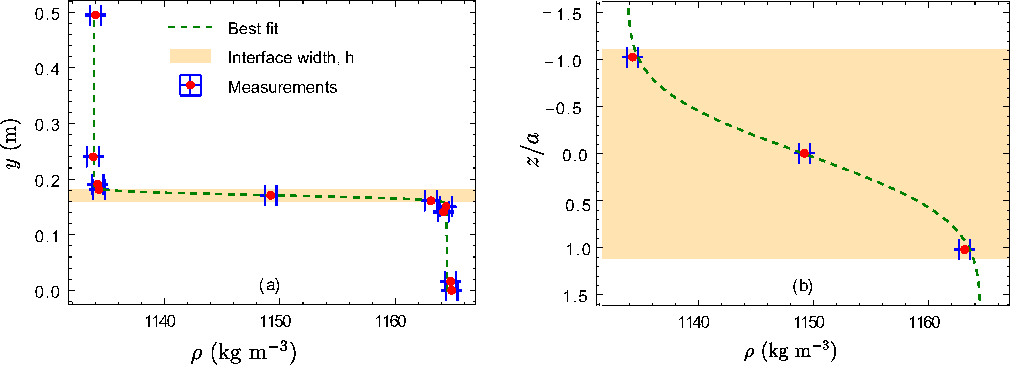}
    \caption{
    (a) A typical vertical density profile, $\rho(y)$ (series B). The shaded area represents the density interface, spanning a 98\% density difference between the upper layer with density $\rho_1$ and the lower layer with density $\rho_2$. The red dots represent the measured density with an uncertainty of $\pm$0.5 kg m$^{-3}$; the green line is the fitted density function for these data points. (b) Zoom into the centered density interface profile, $z=y-h_\text{ref}$, normalized by the sphere diameter, $z/a$, for the region of $\pm 2 z/a$.}
    \label{Figure: Fluid Density Profile}
\end{figure}

The interface thickness value, $h$, is defined to encompass 98\% of the total density variation between upper ($\rho_1$) and lower ($\rho_2$) layers, consistent with previous studies. The observed interface growth by molecular diffusion is approximately $\Delta h \approx 2$~mm over a typical multi-hour experimental session. For the thinnest interfaces in our dataset ($h \approx 10$~mm), this corresponds to a change in the interface-to-sphere ratio of $\Delta(a/h) \approx 20\%$ between first and last spheres in that session. We provide $h$ as a range per series in Table~\ref{Table : Combined Parameters} but treat it as constant within each individual experimental run. Figure~\ref{Figure: Fluid Density Profile} shows a typical measured density profile across the two layers and with a zoom into the interface, where $z = y-h_\text{ref}$ is the vertical position relative to the interface-center height $h_\text{ref}$.

\subsection{Flow Visualization and Particle Tracking}

For flow visualization, we seeded both fluid layers with polyethylene tracers (density 1100 kg m$^{-3}$, diameter 108-180 $\mu$m) as a part of the fluid preparation stage. Despite their small Stokes number, tracers from the upper layer gradually settle into the interfacial region over the term of several hours. This can limit the quality of PIV/PTV experiments over multi-hour sessions due to the reduction in seeding density in the upper layer. On the positive side, this effect provides an additional clear visual indication of the top edge of the interface. 

The imaging system consists of a high-speed FASTCAM SA3 camera (Photron Inc.) with a $1024 \times 1024$ pixel resolution, operating at up to 500 frames per second. The camera captures a $30 \times 30$ cm field of view, providing a spatial resolution of 0.3 mm/pixel. A beam of continuous Nd: YLF laser (6W, 527 nm, MSL Inc.) is shaped into an approximately 0.5 mm-thick light sheet and illuminates the measurement plane.

\subsection{Sphere Release and Tracking}

We developed a computer-controlled release mechanism using a stepper motor and a transparent guide tube. The tube remains submerged to ensure that the spheres begin their descent at nearly terminal velocity. A five-minute interval between releases allows fluid motion to dissipate completely, verified through PIV analysis of tracer particles.

A custom-made image-processing code tracks a sphere at high frame rates to maintain sub-pixel positional accuracy. The sphere's position typically changes less than one pixel between frames, yielding a position uncertainty of approximately 0.3 mm ($\approx 3$\% relative error for 10 mm spheres).

Velocity calculations use a least-squares fit through $n$ sphere centroid positions along the trajectory, following the approach of~\cite{Srdic-Mitrovic1999}. The uncertainty in velocity measurements is given by:
\begin{equation} 
\Delta U=\left[ \frac{12}{n(n+1)(n+2)} \right] ^{\frac{1}{2}} \frac {\Delta y} {\Delta t} \frac {w} {480} \left( \si{m s^{-1}} \right)
\end{equation}
where $\Delta t$ represents the inter-frame time interval, $\Delta y$ denotes the position measurement accuracy in pixels, and $w$ indicates the vertical frame dimension in meters. For our experimental parameters ($\Delta t =1/60 - 1/500$ s, $\Delta y = 1$ pixel, $w = 0.3$ m, 12 and 480 are dimensionless constants in the numerical procedure, see \cite{Srdic-Mitrovic1999}), this yields velocity uncertainty $\Delta U = 5 \times 10^{-4}$~m~s$^{-1}$.

\subsection{Experimental protocol}\label{sec:protocol}

In total, 321 spheres were released across all experiments, with approximately 15 spheres per experimental day. Spheres were released one at a time, with a five-minute waiting interval between successive releases. This interval was verified to be sufficient for the fluid to return to quiescence by PIV analysis: tracers were observed to remain still as the next sphere approached the interfacial layer. Of the 321 releases, 309 yielded successful exponential velocity fits and form the analysis dataset. Among these, 33 trajectories exhibited clear upward motion (bouncing, $U < 0$): 18 in Series~A (water--salt) and 15 in Series~B (water--glycerol). A further 270 trajectories showed a sub-terminal minimum without reversal (209 in Series~A, 61 in Series~B), and 6 spheres in Series~A came to a complete stop at the interface. Among the trajectories with local minima, we distinguish below between spheres that slow down significantly (below 20\% of their terminal velocity, with some approaching 1\% of terminal velocity). We demonstrate below that there is no difference between these spheres and the arrested or bounced spheres in terms of sphere/fluid behavior, except for the shorter retention time.

Each sphere crossing displaces a volume of fluid of order $a^3 \sim 1$~cm$^3$, which is negligible compared to the tank volume ($0.15 \times 0.3 \times 0.6$~m $\approx 27$~liters), so the density profile is not significantly disturbed by individual runs. Several PIV/PTV measurements confirmed that the tracers are stagnant before the next sphere is released. 

In most cases, each experimental day consisted of a single experiment with a fixed fluid mixture and multiple sphere releases. The density profile was measured at the start of each session. When conditions were changed, the tank was fully drained, washed, and refilled with a new fluid combination (see Appendix~A.3 in~\cite{Mortenfeld2024} for the detailed experimental procedure protocol). 

\section{Results and Discussion}
\label{Section : Results}
The following results yield several key observations regarding particle behavior at stratified interfaces, particularly regarding the timeline and mechanisms underlying slowdown, arrest, or bouncing, and emphasize the factors that strongly influence the sphere retention time. 

\begin{figure}[!ht]
    \centering
	\includegraphics[width=\textwidth]{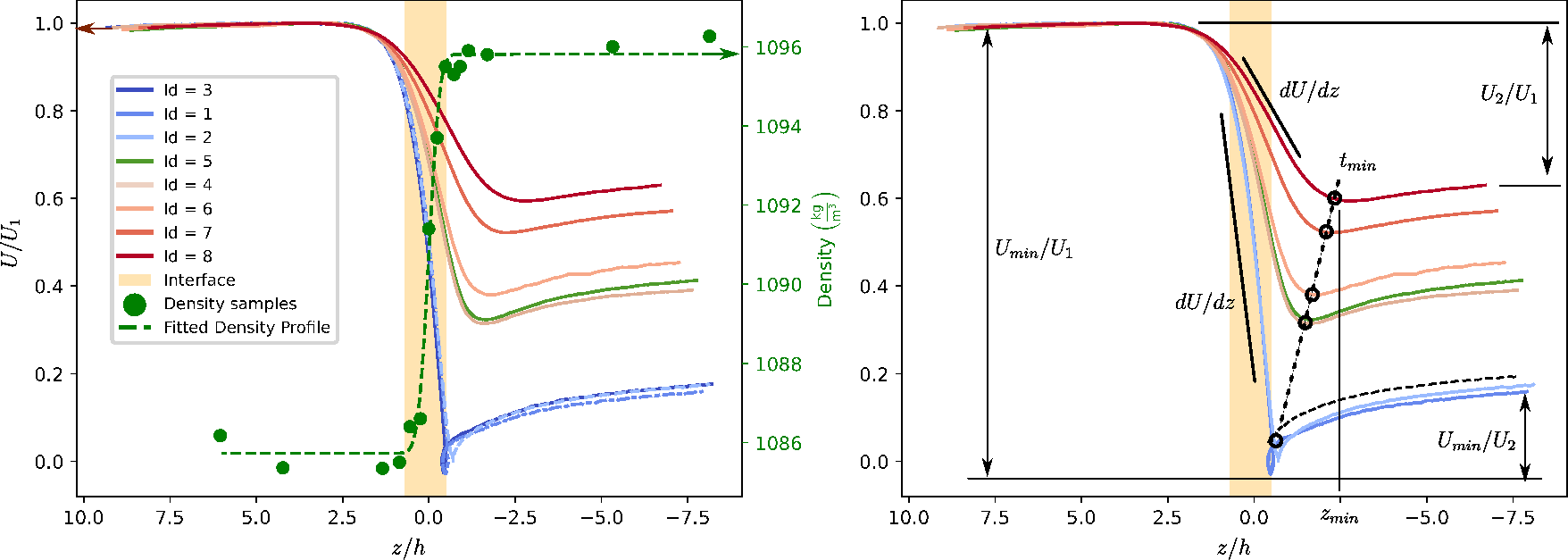}    
    \caption{(left) Archetypic trajectories in the state space of normalized velocity $U(t)/U_1$ versus position $z/h$ (left axis) and density $\rho(z)$ versus position (symbols and dashed line, right axis). (right) Marking of the key observed parameters: ratio of minimal velocity to the terminal velocities, $U_\text{min}/U_{1,2}$, deceleration ratios for bouncing and smooth minima trajectories, $dU/dz$, position of the minimal velocity point in time and relatively to the interface, $t_\text{min}$, $z_\text{min}$, used in our analysis.}
    \label{Figure: four paths}
\end{figure}

We first present a typical set of trajectories in one experimental run in terms of settling velocity, normalized to the terminal velocity in the upper layer $U/U_1$, versus normalized vertical location of the sphere ($z/h$). We also overlay the plot of settling velocity with the shaded region highlighting the interface location and experimental results of density profile for that specific series. We utilize this set to define the archetypes of sphere settling behavior scenarios in \autoref{Figure: four paths}. 

We split the data into five archetypes of sphere trajectories, characterized by their velocity profiles $U(y)$ near the interface, as shown in Figure~\ref{Figure: four paths}:
\begin{enumerate}
\item Monotonic behavior: Spheres that smoothly connect terminal velocity in the upper layer to terminal velocity of the bottom layer (e.g. sphere no. 8 in dark red). Retention time in this case is well defined by the standard equation of motion~\cite{Crowe2011} and for our case it is a fraction of a second. The local minimum of each of these spheres is negligibly small and their terminal velocity ratio $U_2/U_1 \leq 0.7$. This regime can be approximated as motion through an infinite linear stratification regime with similar density gradient of $d \rho/dz \approx \Delta \rho/h$. This is a well understood regime, e.g. \cite{Magnaudet1997}, among others.  
\item Shallow local minimum: Spheres that decelerate to a local minimum velocity below the terminal one with ratios $U_2/U_1 \approx 0.3 - 0.7$, and accelerate to their terminal velocity. Retention time extends to the order of $\sim 5$ seconds. 
\item Deep local minimum: Spheres that decelerate to a very small local minimum velocity $U_2/U_1 \leq 0.25$, and accelerate to their terminal velocity similar to the shallow minimum ones, but much slower (e.g. id=2). Retention time extends to the order of $\sim 10$ seconds. This can be explained by the equation of motion of a sphere with much smaller negative buoyancy (lower net weight) and can only be explained by sphere's low effective density. 
\item Complete stop: Spheres that come to rest ($U_2 \approx 0$) before resuming descent. The retention time is not defined in this case, because some of these spheres could remain in such situation for minutes. In our case all the spheres eventually settled, but we disregarded spheres that floated longer than the camera memory limit to capture their following settling velocity. There is no stopping sphere in this particular run. 
\item Bouncing: Spheres that exhibit temporary upward motion, which in our notation corresponds to a negative settling velocity $U < 0$, before returning to downward settling. In these cases, the retention time increases drastically (up to 10-fold, on the order of 30--150 seconds; see e.g., sphere ID~3).
 
\end{enumerate}

The trends of the dimensional ratios, time, and position of the minimal velocity are clearly emphasized in the right panel—all the ratios of velocities are much smaller for bouncing spheres, and they slow down much closer to the interface at a stronger rate. Some spheres have such a deep minimum but do not bounce or do not bounce for a sufficiently long time to be classified as bouncing with absolute certainty. Yet these spheres seem to have very similar deceleration rates, and we study them together with the bouncing spheres. 

Following these observations, we focus our analysis on finding the scaling and mechanisms that can predict for each sphere and fluid properties its trajectory archetype. We tried different predictive models, including the data-driven approach~\cite{SimonKerenLiron2023}, but without measuring the detailed fluid-sphere coupling, we could not derive a unified view on the timeline of physical events as shown below. 

\subsection{Visual timeline of events leading to bouncing}
\label{subsection:coupling}

\begin{figure*}[tbp]
    \centering
    \includegraphics[width=\textwidth]{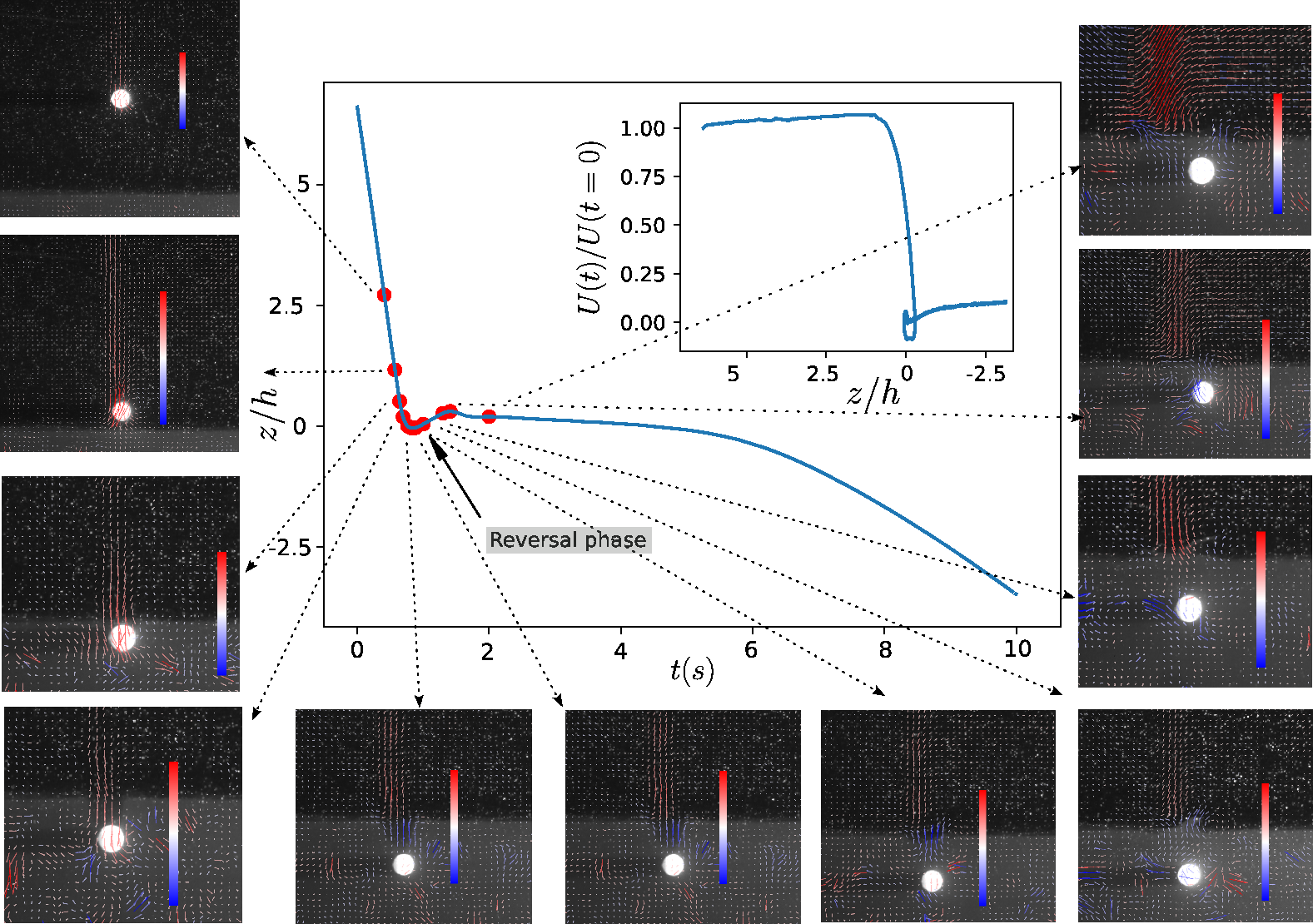}
    \caption{Synchronized analysis of a bouncing sphere trajectory and the surrounding flow field ($Re_1=186$, $Re_2=18$). \textbf{Center:} Sphere position $z/h$ versus time $t\,(s)$. Red dots indicate PIV snapshots. \textbf{Inset:} Phase plot showing velocity reversal in the space of position -- velocity of the sphere. \textbf{Panels:} PIV vector fields (color: vertical velocity component). (1) \textbf{Entrainment:} A persistent wake of light fluid (red vectors point downwards, blue vectors upwards) travels with the sphere. (2) \textbf{Detachment:} The buoyant wake decelerates and detaches from the sphere. (3) \textbf{Reversal:} The returning wake deforms the interface (internal splash); its relaxation (blue vectors) carries the arrested sphere upward.}
    \label{Figure: PIV_Composite}
\end{figure*}

Figure~\ref{Figure: PIV_Composite} presents visually a representative bouncing trajectory as a timeline marked on the various points on the velocity-time plot: the center panel traces sphere position $z/h$ versus time $t$, and the inset phase plot shows the velocity reversal near the interface center.

The PIV vector fields (panels 1--3 in Fig.~\ref{Figure: PIV_Composite}) show three successive stages. (1)~\textit{Entrainment:} during descent, the sphere carries a coherent wake of upper-layer fluid ($\rho_1$) into the denser lower layer; the entrained fluid, buoyant relative to $\rho_2$, decelerates while the sphere continues downward. (2)~\textit{Detachment:} the wake reverses direction and returns toward the interface; simultaneously the sphere decelerates, driven by the increasing ambient density and the removal of the co-moving wake from its trailing side. The PIV measurements at the available spatial resolution do not resolve the separate contributions of these two effects to the total deceleration. (3)~\textit{Internal splash:} the returning wake deforms the interface and excites \BV~oscillations at frequency $N$. A sphere that slowed down to arrest, $U \approx 0$ within this oscillation period is co-moving with the entrained fluid film which provides the transient buoyant restoring force.

The synchronized measurements confirm that during the upward recovery phase, the sphere and its attached fluid film move together. The relative velocity between the sphere and its immediate fluid wake is very small, meaning that the sphere's upward travel is driven by the net buoyancy of this combined, co-moving sphere-film composite. This visual evidence of a coherent, buoyant fluid-solid structure forms the physical basis of our coupled dynamic model.

\clearpage
\subsection{Acceleration analysis}

\begin{figure}
	\centering \includegraphics[width=1\textwidth]{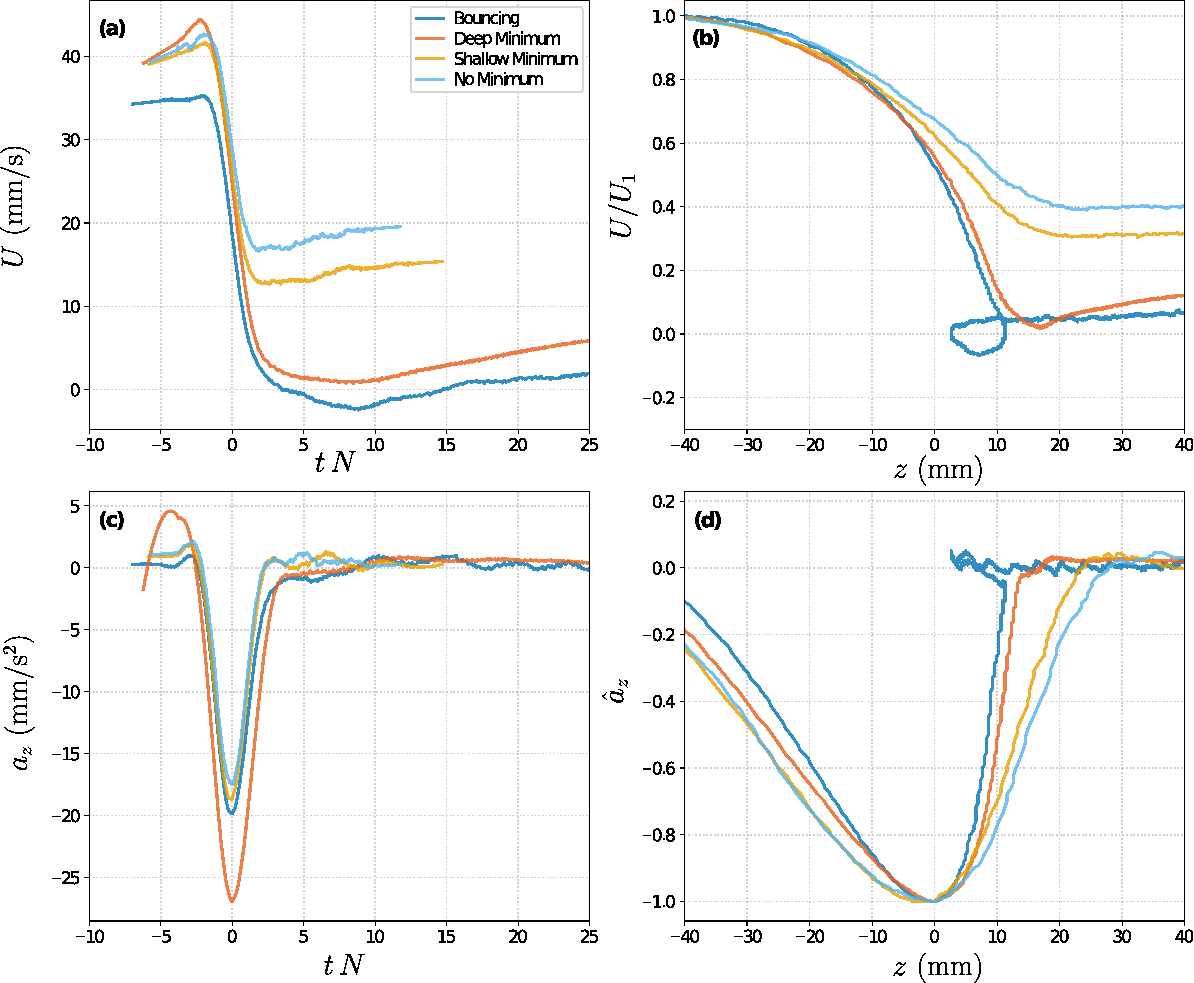}
	\caption{Sphere velocity $U(t)$ (top) and acceleration $a_z = dU/dt$ (bottom) in dimensional (left) and \BV-normalized (right) form for a sample of bouncing trajectories. The collapse of the normalized acceleration curves (panel~c) confirms that all bouncing events share the same stratification-driven mechanism, despite spheres reversing at different depths below the interface (panel~d).}
	\label{fig:velocity_acceleration}
\end{figure}

We proceed with a quantitative analysis of trajectories focusing on the bouncing ones through their settling velocity and acceleration in two modes: as function of time, as seen by the equation of motion, and as a function of vertical position, as it would be in the correct context of the local buoyancy balance. Figure~\ref{fig:velocity_acceleration} shows $U(t)$ and $dU/dt$ for a set of bouncing trajectories in dimensional form (panels a, b) and normalized by $1/N$ relative to $h_\mathrm{ref}$ (panels c, d). The normalized acceleration profiles collapse during the deceleration phase (panel c), showing that all bouncing events share the same stratification-driven deceleration. The velocity passes smoothly through zero (panel d), with no abrupt change of sign that would indicate an impulsive force from the wake converting to jet.

\begin{figure}
	\centering 
	\includegraphics[width=\textwidth]{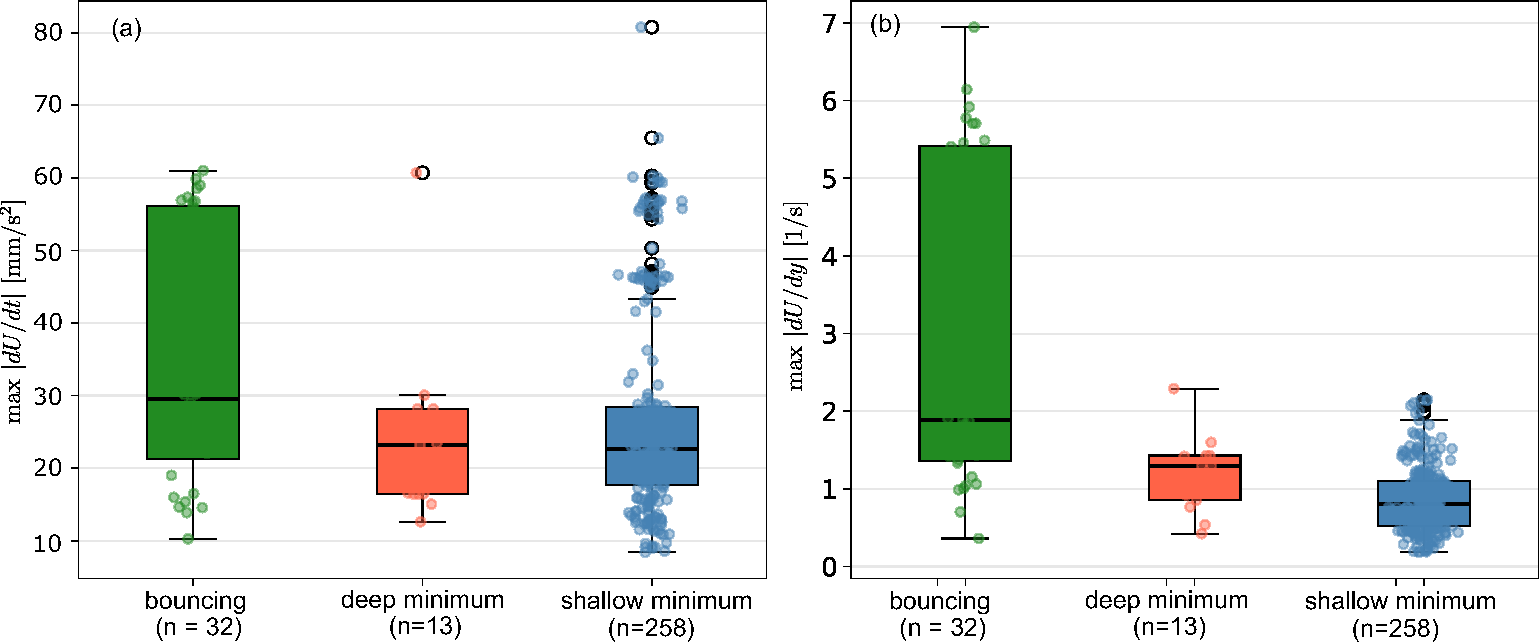}
	\caption{Statistics of acceleration values during crossing in terms of $dU/dt$ and $dU/dy$ }
	\label{fig:dudt_dudy}
\end{figure}

A more obvious difference between the trajectory archetypes lies in their deceleration profiles with respect to the fluid density gradient. The four different types of trajectories, classified by their behavior after crossing (shallow/deep minima, bouncing, or stopped), are clearly sorted by the spatial rate of deceleration, $dU/dy = (dU/dt)/U$. Following this observation, it is instructive to check the statistics of acceleration values as $dU/dt$ and $dU/dy$, presented as boxplots in Fig.~\ref{fig:dudt_dudy}. The right panel showing $dU/dy$ is on a log scale emphasizing that this is the dominant effect that determines sphere trajectory archetypes or ``sorts'' the spheres between bouncing, deep, and shallow minimum. We will rely on this observation in the following construction of our dynamic coupled model. 

\subsection{How viscosity affects comparison with state-of-the-art scaling}

\begin{figure}[tbp]
    \centering
    \includegraphics[width=\textwidth]{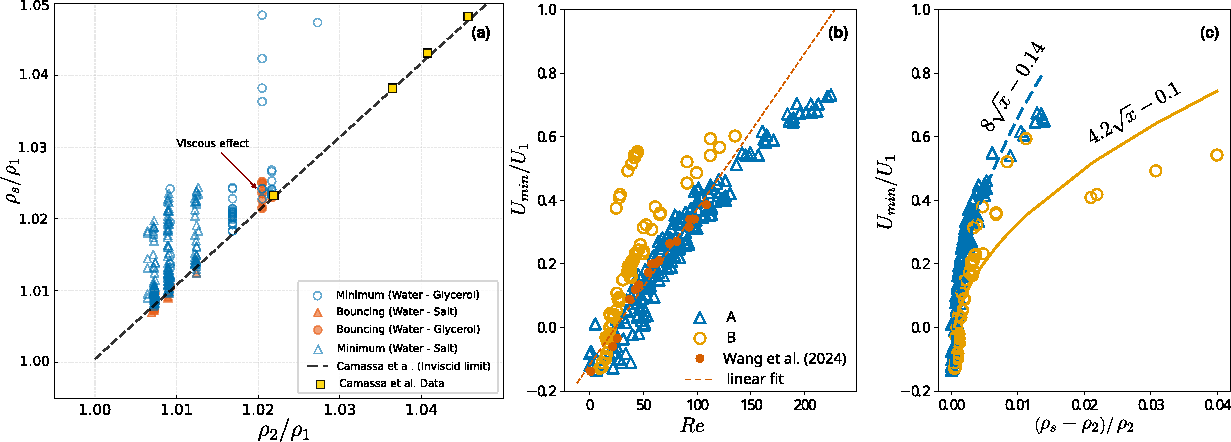}
    \caption{(a) Density triplets $\rho_{s}/\rho_{1}$ versus $\rho_{2} / \rho_{1}$. The black and red dashed lines are the critical density ratios according to Eq.~\eqref{eq:critical_density}. Each ``column'' of symbols is from a different experiment date with $\rho_{1}$ and $\rho_{2}$ fixed and $\rho_{s}$ vary per sphere. In squares, we add data of stopping particles from Table 1 in the original paper~\cite{Camassa2022}. (b) Minimum velocity $U_\text{min}$ and digitized results of Wang et al.~\cite{Wang2024} according to the scaling in Eq.~\eqref{eq:U_min}. (b) $U_\mathrm{min}/U_1$ vs $Re_2$ and (c) vs $\Delta \rho_2/\rho_2$.}
    \label{Figure: cammasa plot}
\end{figure}

We compare our results with the experiments of Camassa et al.~\cite{Camassa2022} in \autoref{Figure: cammasa plot}a in the same form as the authors proposed to support the critical density triplet prediction. Our experiments are conducted in a slightly different parameter range of density jumps and particle diameters, closer to the work of Wang et al.~\cite{Wang2024}, but with the same interface sphere-to-width ratio, $a/h \approx 0.2 - 0.5$. Data from Series A (water-salt stratification) are represented by triangles, and data from Series B (water-glycerol stratification) are represented by circles. The square symbols are from the Camassa et al.~\cite{Camassa2022} table of parameters for the particles that stopped in the interface (the authors in \cite{Camassa2022} did not report on bouncing spheres). The line is Eq.~\eqref{eq:critical_density} with constants $a_{1,2}$ fitted to the experimental results of~\cite{Camassa2022}.

Series A results fall relatively close to the critical density triplet prediction linear fit, Eq.~\eqref{eq:critical_density}. However, for Series B, with a higher viscosity ratio, the results are above the predicted line, meaning that much denser spheres well above the critical line can still bounce. 

Similarly, we expected to match the scaling proposed by Wang et al.~\cite{Wang2024}, shown in \autoref{Figure: cammasa plot}b-c. We first test the Reynolds-based scaling of the minimum velocity shown by the authors (see Figure 19 in \cite{Wang2024}) in \autoref{Figure: cammasa plot}b, including the negative velocity values associated with bouncing. We obtain a reasonably good fit for the slower spheres ($Re_2 < 100$) in Series A, as shown in \autoref{Figure: cammasa plot}b, but both Series B and the faster spheres in Series A show deviation from that scaling. We understand that the linear scaling of $U_\text{min}/U_1$ does not represent the full range of experiments, either for the faster spheres or for Series B. 

Similarly, we test Wang et al.~\cite{Wang2024} expression that scales this minimum velocity with the density difference as a leading order mechanism, following theoretical framework of critical density triplets of \cite{Camassa2022a}: 
\begin{equation}\label{eq:delta_rho2}
    \Delta\rho_2 = \frac{3C_{d,2}v_2^2 Re_2^2}{4ga^3},
\end{equation}
The authors \cite{Wang2024} obtained the square root relation between the minimum velocity $U_\mathrm{min}$ and the density jump $\Delta \rho$ as:
\begin{equation}\label{eq:U_min}
    \frac{U_\mathrm{min}}{U_1} = c_1 \sqrt{\frac{4ga^3}{3\nu_2^2 C_{d,2}}} \Delta\rho_2^{1/2} + c_2.
\end{equation}

Also in this case, Series A replicates the findings of~\cite{Wang2024}, as shown in \autoref{Figure: cammasa plot}c with symbols for experimental data and the proposed scaling of Eq.~\eqref{eq:U_min}. In this plot, the attempt to scale both series (A and B) simultaneously fails.

While the scaling form appears qualitatively correct and both $\Delta\rho_2$ and $U_\text{min}/U_1$ relate to the density excess to the power of 0.5, the failure of a single scaling set based solely on density~\cite{Camassa2022, Wang2024} for Series B implies that the upper-layer viscosity is a crucial physical parameter. We hypothesize that higher viscosity allows the boundary layer of crossing spheres to entrain a larger volume of lighter, buoyant upper-layer fluid ($\rho_1$) deep into the denser lower-layer fluid. The next section formalizes this entrained-film mechanism in a phenomenological dynamic coupled model using viscosity-dominated boundary-layer scaling arguments.

\subsection{Dynamic model: coupled equations of motion of the sphere-film system}
\label{subsection:first_principles}

To capture the physical mechanisms of the crossing, we formalize the visual evidence from our synchronized PIV/PTV measurements (Section~\ref{subsection:coupling}) into a phenomenological dynamic model. The observations demonstrate that the entrained fluid does not shed instantly but forms a co-moving buoyant envelope. By treating the sphere and this envelope as a single Lagrangian composite body, we extend the entrained-fluid framework of Verso et al.~\cite{Verso2019} to regimes involving deep minima and bouncing.

The state of the composite system is defined by the sphere's position $y$, downward velocity $U=\dot{y}$, and the volume of the entrained upper-layer fluid $V_e(t)$. Their evolution is governed by coupled momentum and entrainment balances:
\begin{equation}\label{eq:momentum_ode}
  \bigl(\rho_s\,V_s + \rho_1\,V_e\bigr)\,\dot{U}
  = \underbrace{\bigl(\rho_s - \rho_f\bigr)\,V_s\,g}_{\text{net weight}}
  - \underbrace{\bigl(\rho_f - \rho_1\bigr)\,V_e\,g}_{\text{entrained buoyancy}}
  - \underbrace{\tfrac{1}{2}\,\rho_f\,A_s\,C_d\,U|U|}_{\text{drag}},
\end{equation}
\begin{equation}\label{eq:entrainment_ode}
  \dot{V}_e
  = \underbrace{\tilde{\beta}\,\frac{V_s}{\rho_2 - \rho_1}\,U\,
    \left|\frac{d\rho_f}{dy}\right|}_{\text{entrainment }(U>0)}
  - \underbrace{\alpha_d\,V_e}_{\text{viscous drainage}},
\end{equation}
where $V_s = \pi a^3/6$ is the sphere volume and $A_s = \pi a^2/4$ its cross-sectional area. 

In Eq.~\eqref{eq:momentum_ode}, the inertial mass includes both the sphere and the entrained fluid ($\rho_1 V_e$), whose buoyancy opposes the net weight. Because the relative velocity between the sphere and its envelope is negligible, the standard drag force acts on the composite system. In Eq.~\eqref{eq:entrainment_ode}, the entrained volume grows during descent ($U > 0$) across the density gradient with an efficiency parameterized by $\tilde{\beta}$, and drains continuously due to gravity and viscosity at a rate $\alpha_d$.

This formulation relies entirely on mass and momentum conservation. Although the core framework has shown consistency with numerical simulations and historical data~\cite{Srdic-Mitrovic1999, Verso2019, Boetti2022}, applying it directly to the arrest and bouncing thresholds reveals the true structural role of the entrained fluid film.

\subsection{Analytical limits and trajectory archetypes}\label{subsec:application}

\begin{figure}[tbp]
  \centering
  \includegraphics[width=\textwidth]{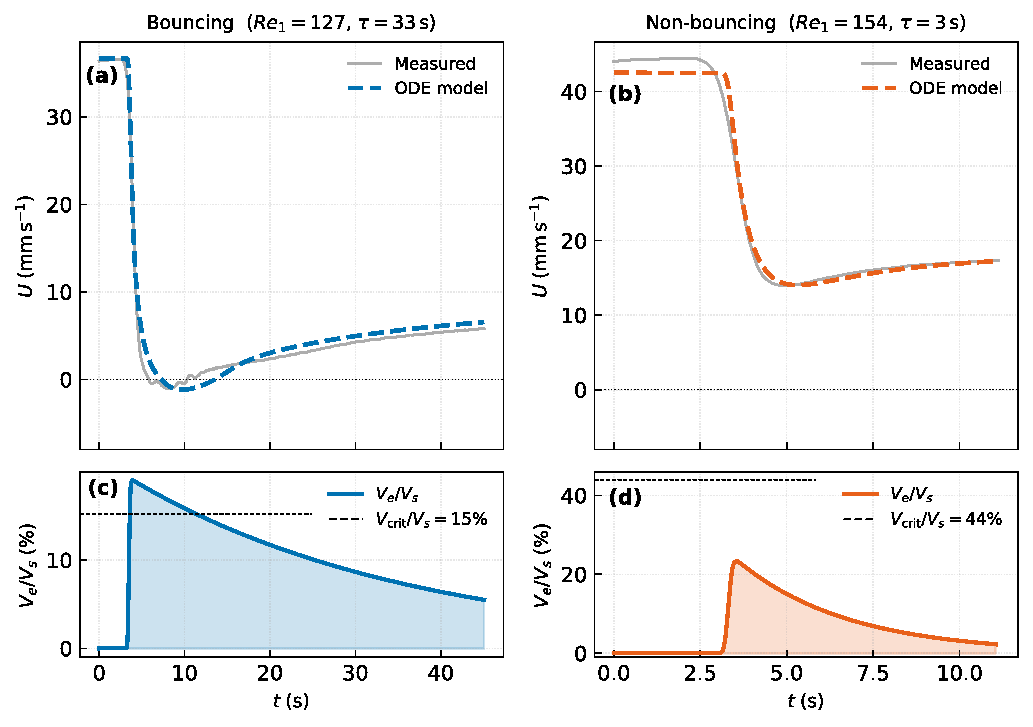}
  \caption{ODE model versus measured trajectories. Solid gray line: measured sphere speed $U(t)$; dashed colored line: ODE integration of Eqs.~\eqref{eq:momentum_ode}--\eqref{eq:entrainment_ode}. (a)~Bouncing trajectory ($Re_1 = 127$): the sphere decelerates to arrest and reverses; the model reproduces the turnaround timing and recovery. (b)~Non-bouncing trajectory ($Re_1 = 154$): the sphere decelerates to a sub-terminal minimum and recovers; the model reproduces minimum speed and recovery timescale. (c,d)~Entrained volume $V_e(t)/V_s$ for the same two trajectories. Dashed horizontal line: $V_\mathrm{crit}/V_s = (\rho_s-\rho_2)/(\rho_2-\rho_1)$. In the bouncing case the model-inferred $V_e$ exceeds $V_\mathrm{crit}$; in the non-bouncing case it does not.}
  \label{fig:ode_comparison}
\end{figure}

This physically plausible formulation replaces the empirical `virtual spring' of Abaid
et al.~\cite{Abaid2004} with a rational, mass-conserving physical mechanism. Numerically integrating this ODE system successfully reproduces the different trajectory archetypes (\autoref{fig:ode_comparison}). The model captures both the velocity reversal of a bouncing sphere and the smooth deceleration of a non-bouncing sphere. Furthermore, the model shows clearly how $V_e$ grows above the critical value set by the density differences for the bouncing case, emphasizing that ODE solves a dimensionally correct and physically sound system of equations. 

A key insight emerges from evaluating the momentum equation at its stationary points. With Edwards drag $C_d \propto Re^{-1/2}$, the drag force scales as $U^{3/2}$. Setting $\dot{U}=0$ in Eq.~\eqref{eq:momentum_ode} gives three physical limits: upper terminal velocity ($U_1$) above the interface, where $V_e=0$, drag balances net weight in the upper fluid, lower terminal velocity ($U_2$), far below the interface after the film has drained ($V_e=0$), and velocity minimum ($U_\mathrm{min}$), because in the lower fluid, the buoyant film ($V_e>0$) retards the sphere to $U=0, dU/dt=0$. Normalizing by the $U_2$ balance isolates the film's effect, marked by the dashed line in \autoref{fig:ode_comparison}:
    \begin{equation}\label{eq:Umin_formula}
      \left(\frac{U_\mathrm{min}}{U_2}\right)^{3/2} = 1 - \frac{V_e/V_s}{\varphi_c},
      \qquad \varphi_c = \frac{\rho_s-\rho_2}{\rho_2-\rho_1}.
    \end{equation}

We understand that to slow down the sphere to zero velocity, ($U=0$, $\dot{U} < 0$), the supply of entrained volume, the entrained buoyancy should exceed the net weight:
\begin{equation}\label{eq:bounce_crit}
 \frac{V_e}{V_s} > \varphi_c.
\end{equation}

It is noteworthy that while $\varphi_c$ mirrors the critical density threshold of Camassa et al.~\cite{Camassa2022}, here it acts as a dynamic constraint: heavy spheres (heavier that set by the critical density triplet) can bounce if their boundary layer entrains a sufficient buoyant volume $V_e > \varphi_c V_s$.

\subsection{Bulk classifier from the ODE: the velocity-ratio parameter $\Pi$}
\label{subsection:froude_mapping}

The ratio of the terminal settling velocities $U_2/U_1$ encodes the ratio of net driving forces, corrected for viscosity through the drag law. Raising this ratio to the drag-law exponent ($3/2$) defines the dimensionless bulk parameter $\Pi$:
\begin{equation}\label{eq:Pi_definition}
  \Pi \equiv \left(\frac{U_2}{U_1}\right)^{3/2}
  = \frac{\rho_s-\rho_2}{\rho_s-\rho_1}\left(\frac{\nu_1}{\nu_2}\right)^{1/2}.
\end{equation}
$\Pi$ classifies the trajectory archetype from bulk properties alone. Using Eq.~\eqref{eq:Umin_formula} and the identity $U_\mathrm{min}/U_1 = (U_\mathrm{min}/U_2)(U_2/U_1)$, we normalize the minimum velocity to establish a zero-parameter separator:
\begin{equation}\label{eq:separator_corrected}
  \frac{U_\mathrm{min}/U_1}{\Pi^{2/3}} = \frac{U_\mathrm{min}}{U_2} = \left(1 - \frac{V_e/V_s}{\varphi_c}\right)^{2/3}.
\end{equation}
This separator demonstrates that the natural zero-film limit is exactly 1. Trajectories entraining negligible film ($V_e \approx 0$) approach this upper bound, while those deploying a significant buoyant parachute ($V_e > 0$) strictly fall below 1.

\subsection{Predictability and experimental validation}
\label{subsection:predictability}

\begin{figure}
  \centering
  \includegraphics[width=\textwidth]{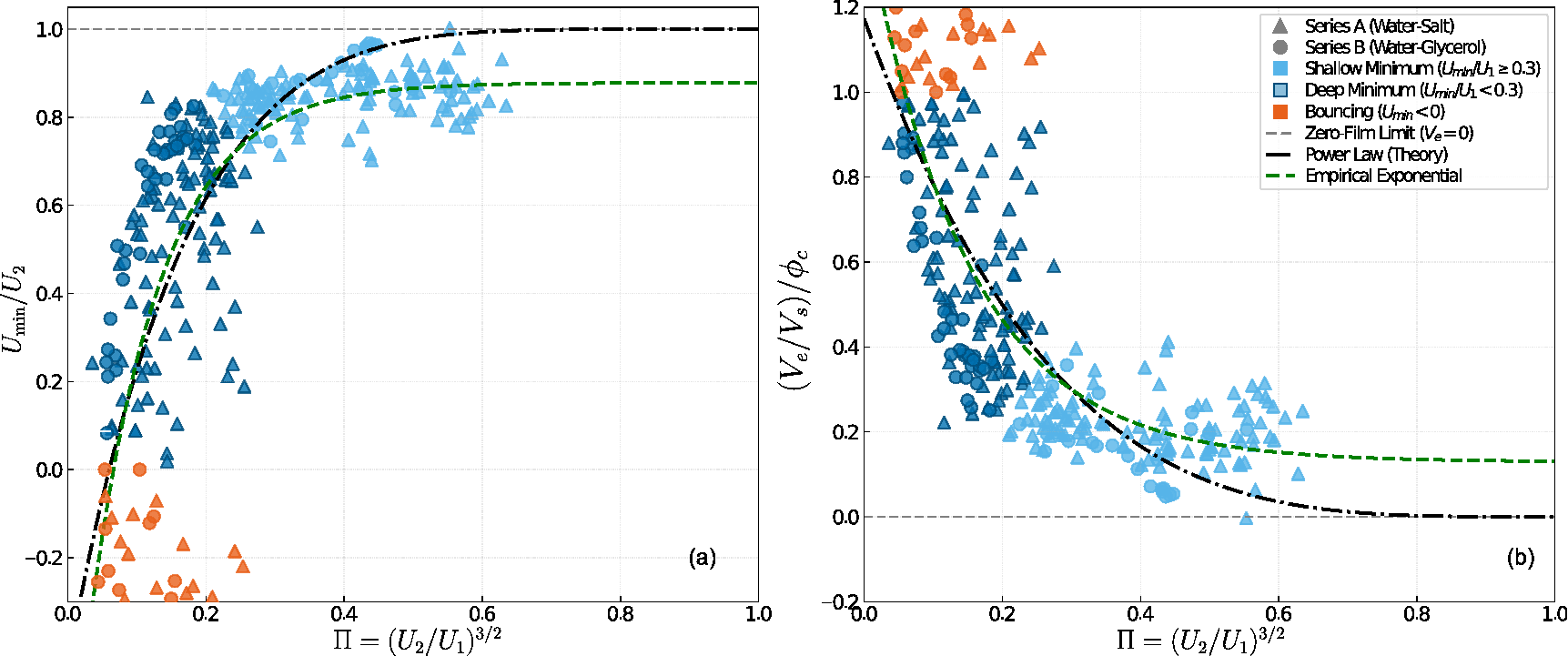}
  \caption{(a) Normalized minimum velocity $U_\mathrm{min}/U_2$ versus $\Pi = (U_2/U_1)^{3/2}$. The dashed line at 1.0 represents the absolute theoretical limit for a sphere that entrains no wake ($V_e = 0$). (b) Parachute Strength $(V_e/V_s)/\phi_c$ versus $\Pi$, isolating the volume of the entrained fluid film. As $\Pi \to 0$, the entrained wake volume grows to exceed the critical buoyancy threshold ($\phi_c$). Two possible fits (a theoretical power law and an empirical exponential) unify the prediction for shallow, deep minimum, and bouncing spheres.}
  \label{fig:separator}
\end{figure}

\autoref{fig:separator} plots the normalized minimum velocity $U_\mathrm{min}/U_2$ against the dimensionless bulk coordinate $\Pi$. The dashed line at exactly 1.0 represents the absolute theoretical limit for a sphere that entrains no wake ($V_e = 0$). By correctly scaling the axes, the physical bifurcation becomes starkly apparent: trajectories that cross smoothly approach the zero-film limit ($U_\mathrm{min}/U_2 \to 1$), while any sphere that deploys a ``buoyant parachute'' ($V_e > 0$) is dynamically forced below this threshold. Both the water-salt and water-glycerol series collapse perfectly under this bounding limit, confirming that viscosity is correctly absorbed by the $\Pi$ parameter.

\subsection{Spatial scaling: the penetration depth ($z_{\min}$)}
\label{subsection:spatial_scaling}

While the velocity ratio parameter $\Pi$ governs the severity of the deceleration, the physical depth at which this minimum occurs, $z_{\min}$, is governed by the sphere's inertia and the spatial decay of the entrained wake. Transforming the temporal wake detachment equation from our ODEs into the spatial domain introduces a characteristic detachment length, $L_{\mathrm{detach}} \propto \bar{U}/\alpha_d$. This represents the vertical distance the sphere must travel for its entrained wake volume to decay significantly. 

For light, buoyancy-dominated spheres (bouncing regime), the initial kinetic energy is small. Their braking distance is so short compared to their detachment length ($z_{\min} \ll L_{\mathrm{detach}}$) that the wake does not significantly detach during the initial penetration. In this limit, they arrest rapidly near the interface as if decelerating against a constant buoyant force. 

Conversely, dense, inertia-dominated spheres (smooth crossing regime) enter the lower layer with massive kinetic energy. They require a long braking distance during which the wake actively detaches. This continuous loss of entrained volume weakens the buoyant retarding force, causing them to penetrate much deeper into the lower fluid before reaching their minimum velocity. This interplay between inertia and the characteristic detachment length explains the stark differences in $z_{\min}$ across the trajectory archetypes.

\subsection{Retention time prediction}

\begin{figure}
  \centering
  \includegraphics[width=0.9\textwidth]{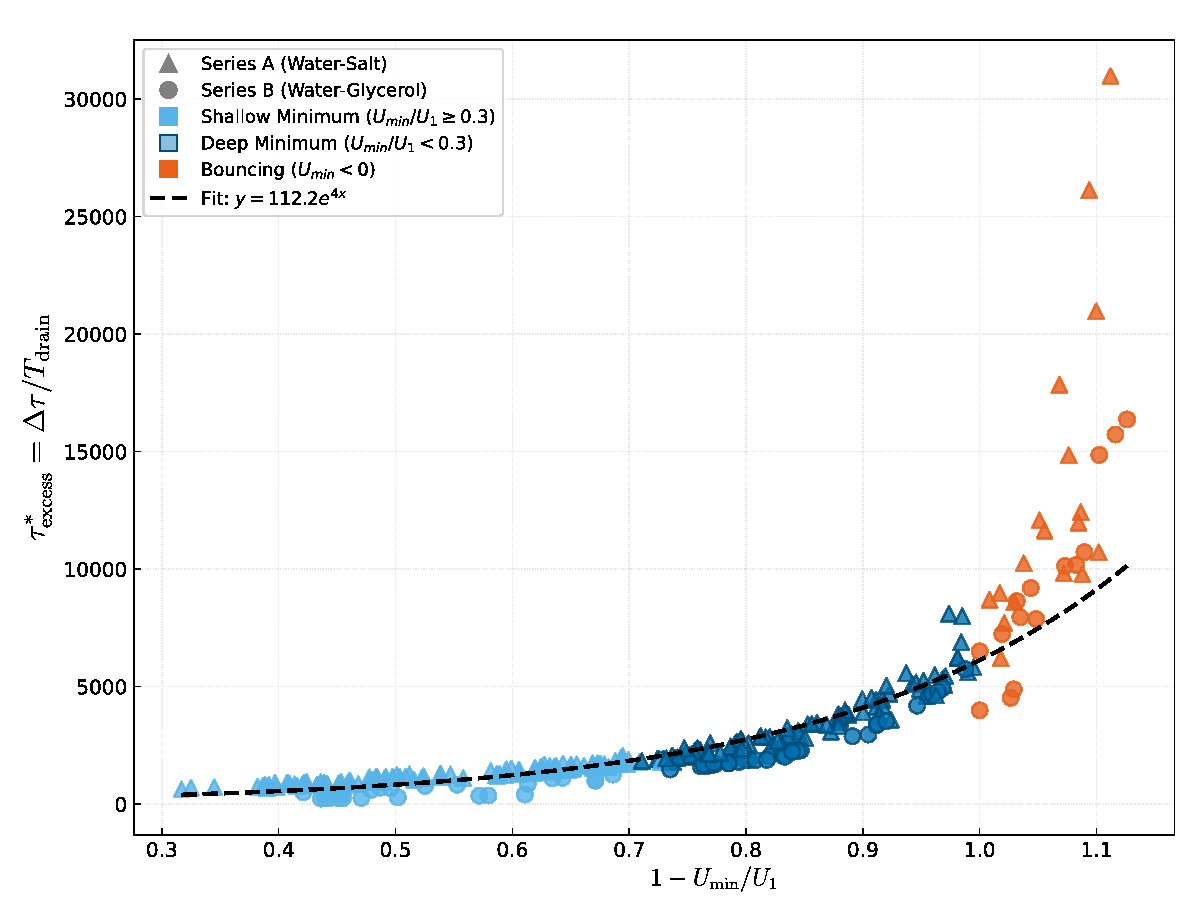}
  \caption{Retention time master curve. The dimensionless excess retention time ($\tau_{\mathrm{excess}}^* = (\tau_\mathrm{meas} - \tau_\mathrm{smooth}) / T_{\mathrm{drain}}$) is plotted against the deceleration severity ($1 - U_\mathrm{min}/U_1$). As severity approaches 1.0 (deep minima and bouncing spheres), the dimensionless retention time grows. Dividing by the $T_{\mathrm{drain}}$ scale precisely collapses the high-viscosity water-glycerol data onto the same fundamental curve as the low-viscosity water-salt data, providing a unified predictive scaling for interface retention across different viscosity regimes.}
  \label{fig:retention_time}
\end{figure}

\autoref{fig:retention_time} shows measured retention times versus the analytical
smooth-crossing baseline $\tau_\mathrm{smooth}=3h/(U_1+\sqrt{U_1 U_2}+U_2)$ for all trajectories. This baseline is
derived by integrating the ODE with $V_e=0$ throughout: the sphere crosses a linear
density gradient of width~$h$ in quasi-static descent at the local terminal speed
$U(y)^{3/2}=U_1^{3/2}[1-(1-\Pi)y/h]$, giving an exact closed-form integral. It
represents the shortest possible interface residence time, the time for a sphere with no entrained film to cross from $U_1$ to $U_2$. For the limiting cases: $U_1=U_2=U$ gives $\tau_\mathrm{smooth}=h/U$ (uniform fluid); $U_2\to0$ gives
$\tau_\mathrm{smooth}=3h/U_1$.

All measured trajectories exceed $\tau_\mathrm{smooth}$ because the entrained buoyant film retards the sphere's descent. The excess retention time, $\Delta \tau = \tau_\mathrm{meas} - \tau_\mathrm{smooth}$, represents the time the sphere spends clearing its wake.

For bouncing spheres starting from zero velocity, this recovery is governed not by an inertial advection effect, but by the buoyancy-driven viscous-gravitational drainage of the residual film. By dimensional analysis, the fundamental timescale for thin-film viscous drainage over a sphere of radius $a$ is:
\begin{equation}
    T_{\mathrm{drain}} = \frac{\rho_2 \nu_2}{\Delta \rho \, g \, a}.
\end{equation}
This drainage timescale explicitly isolates the additional role of the lower-layer viscosity $\nu_2$. While raw retention times in the glycerol system (Series~B) are two to eight times longer than those in the salt system (Series~A), normalizing the excess retention time by this fundamental timescale ($\tau_{\mathrm{excess}}^* = \Delta \tau / T_{\mathrm{drain}}$) collapses the data across both fluid systems.

This provides an applicative recipe: the retention time of an inertial sphere crossing a stratified interface can be predicted from bulk fluid properties by estimating $\tau_\mathrm{meas} \approx \tau_\mathrm{smooth} + C \cdot T_{\mathrm{drain}}$, bypassing the need for complex numerical integration or empirical spring constants. We present this empirical fit in \autoref{fig:retention_time}.

\section{Conclusions}\label{Section: Conclusions}

The bouncing of inertial spheres at density interfaces is fundamentally the dynamic response of a coupled solid-fluid system. As a sphere crosses an interface, it entrains a buoyant boundary layer of lighter fluid, creating a transient composite body whose net buoyancy can temporarily overcome gravity and reverse the sphere's descent. Synchronized PIV/PTV measurements across multiple trajectories in both water-salt and water-glycerol systems provide direct visual confirmation of this mechanism, showing that the sphere and its buoyant envelope move together with near-zero relative velocity during the deceleration and recovery phases.

By formalizing momentum and mass conservation for this composite body in a phenomenological dynamic model, we replace previous empirical and static-density frameworks with a predictive physical mechanism. Evaluating this model at its stationary points reveals a universal, zero-parameter phase diagram that governs all trajectory archetypes. We show that the normalized minimum velocity ($U_{\mathrm{min}}/U_2$) must strictly approach $1.0$ for spheres with negligible wake, whereas spheres deploying a significant buoyant parachute are dynamically forced below this limit. This separator elegantly classifies smooth crossings, deep minima, and bouncing trajectories across drastically different fluid systems.

Crucially, this framework highlights the dual role of viscosity, which density-only models overlook: a higher upper-layer viscosity kinematically thickens the boundary layer to entrain more fluid, while the lower-layer viscosity dynamically dictates the resisting drag. Furthermore, transforming the model into the spatial domain explains the physical depth of the velocity minimum, revealing a bifurcation between inertia-dominated spheres that plunge deeply and buoyancy-dominated spheres that arrest exactly at the interface.

A direct consequence of this buoyant parachute mechanism is a sharp bifurcation in retention time. The entrained film sheds through buoyancy-driven viscous-gravitational drainage. By normalizing the measured retention times against a fundamental thin-film drainage timescale, $T_{\mathrm{drain}}$, we achieve a unified collapse of both the low-viscosity water-salt and high-viscosity water-glycerol datasets onto a single master curve.

Several questions remain open. First, the entrained film volume $V_e$ is currently inferred from the measured velocity minimum; an independent \textit{a priori} estimate of $V_e$ from boundary-layer theory or direct film-thickness measurement would close this gap and yield a fully predictive model. The drainage timescale is established here as a rigorous scaling relationship, but predicting its absolute value requires knowing the residual film thickness, which was not directly measured. Second, our findings apply specifically to the sharp-interface regime; extending the framework to thicker interfaces, non-spherical particles, or slowly varying density gradients would broaden its applicability. Ultimately, these physical insights provide a robust foundation for understanding and predicting retention times in industrial settling tanks, marine snow formation, and microplastic trapping at haloclines.

\section*{Acknowledgments}
The authors thank Avraham Balas for his assistance with the experimental setup and the design of the sphere-release mechanism. We also thank the ``ADAMA Center for Novel Delivery Systems in Crop Protection'' at Tel Aviv University for providing access to precision density measurement equipment.

\section*{Declarations}
\subsection*{Funding}
This research was supported by the Israel Science Foundation (grant number 441/2) and the Gordon Center for Renewable Energy at Tel Aviv University.

\subsection*{Data availability}
The data supporting the conclusions of this study are presented in the figures, tables, and appendices of the manuscript. Raw sphere trajectory data (velocity vs.\ time for all 309 trajectories) and the experimental parameter table are available from the corresponding author upon reasonable request.

\subsection*{Conflicts of interest}
The authors declare no financial or proprietary interests in any materials discussed in this article.

\subsection*{AI usage disclaimer}
During the preparation of this work, the authors used generative AI tools (Claude, Gemini) for text editing and preparation of Python codes for data analysis. The authors reviewed and edited the output as needed and take full responsibility for the content of the published article.

\bibliographystyle{elsarticle-num}

\clearpage
\appendix

\section{Detailed description of raw data analysis}
\label{Appendix:RecoveryFit}

There are several basic properties that need to be extracted from the PTV of a moving sphere before the theoretical framework can be applied from first principles. First, there is an open debate about the $C_d$ correlation, which is empirical rather than theoretical for finite-Reynolds-number spheres. Second, the uncertainty of sphere diameter is much lower compared to that of the sphere density; since it is extremely difficult to reduce this density uncertainty by direct weighing, the alternative is to apply falling-sphere analysis to the unsteady motion of the sphere. We present in this appendix all the details of the raw data analysis to support our findings. The important measurements are $\rho_s$, $C_d$, recovery time in the bottom layer, $\tau$, along with the terminal velocity in both layers, $U_{2,\infty}$ and $U_{1,\infty}$, all cross-validated across all trajectories and compared to the theoretical results. 

Most trajectories in Series A and B do not reach their terminal velocity in the bottom layer, $U_{2}$, as used in the theoretical predictions.

After a sphere crosses the density interface and enters the denser lower layer, its velocity recovers from its post-crossing minimum toward the lower-layer terminal velocity~$U_2$. During recovery, the entrained upper-layer film sheds diffusively while drag accelerates the sphere. In this appendix we present a simple analytical model for this recovery, show how it yields both the recovery timescale~$\tau$ and the sphere density~$\rho_s$, and describe the calibration of the drag prefactor~$\alpha$ against eight spheres of independently measured density.

\subsection{Exponential recovery model}

We model the recovery-phase speed as an exponential approach to the lower-layer terminal velocity:
\begin{equation}\label{eq:exp_recovery}
	U(t) = U_\infty \left[1 - \exp\!\left(-\frac{t - t_0}{\tau}\right)\right], \qquad t \geq t_0,
\end{equation}
where $U_\infty$ is the asymptotic terminal speed in the lower layer, $t_0$ is a virtual origin (the time at which the extrapolated velocity passes through zero), and $\tau$ is the recovery timescale. The three parameters $(U_\infty,\, t_0,\, \tau)$ are determined by nonlinear least-squares fitting of Eq.~\eqref{eq:exp_recovery} to the measured speed in the recovery window, defined as the interval from the post-crossing velocity minimum to the end of the recorded trajectory.

\subsection{Deriving sphere density from the fit}

Once $U_\infty$ is known, the sphere density is obtained from the lower-layer terminal velocity balance. At terminal velocity the drag force equals the net weight:
\begin{equation}\label{eq:drag_balance}
	(\rho_s - \rho_2)\,V_s\,g = \tfrac{1}{2}\,\rho_2\,A_s\,C_d\,U_\infty^2,
\end{equation}
with $C_d = \alpha\,Re^{-1/2}$, where $Re = U_\infty\,a / \nu_2$ and $\alpha$ is a universal drag prefactor valid in the intermediate Reynolds-number regime of the experiments ($Re \sim 1$--$200$). Solving for~$\rho_s$:
\begin{equation}\label{eq:rho_from_Uinf}
	\rho_s = \rho_2 + \frac{3\,\alpha\,\rho_2\,\sqrt{\nu_2}\,U_\infty^{3/2}}{4\,g\,a^{3/2}}.
\end{equation}
This inversion replaces the need for independent density measurements and reduces the free parameters of the trajectory model from four (in the earlier ODE-based approach) to three: $U_\infty$, $t_0$, and~$\tau$.

\subsection{Physical meaning of the recovery timescale \texorpdfstring{$\tau$}{tau}}

The fitted timescale~$\tau$ combines two concurrent relaxation processes: (i)~drag-driven acceleration of the sphere toward~$U_2$ with a linearized timescale $\tau_\mathrm{drag} = (\rho_s V_s + \rho_1 V_e) / (\partial F_\mathrm{drag}/\partial U)$, and (ii)~diffusive shedding of the entrained lighter film. Because both processes are approximately exponential, they contribute additively to the overall recovery timescale.

The key finding is that $\tau$ sharply discriminates between the two settling states:
\begin{itemize}
	\item \textbf{Bouncing spheres}: $\tau \approx 18$--$33$~s (mean $26$~s), reflecting a large entrained film volume that takes tens of seconds to shed.
	\item \textbf{Minimum (non-bouncing) spheres}: $\tau \approx 3$--$4$~s (mean $3.4$~s), consistent with a thin boundary-layer film that sheds rapidly.
\end{itemize}
The ratio $\tau_\mathrm{bouncing}/\tau_\mathrm{minimum} \approx 8$ reflects the fundamentally different film volumes entrained in the two regimes (\autoref{fig:tau_timescale}). This timescale constitutes one component of the total retention time~$\tau_r$ defined in Section~3.4: the recovery portion of the trajectory from the velocity minimum back to terminal speed accounts for the majority of $\tau_r$ in bouncing spheres.

\subsection{Drag law validation}

\begin{figure}[tbp]
	\centering
	\includegraphics[width=\textwidth]{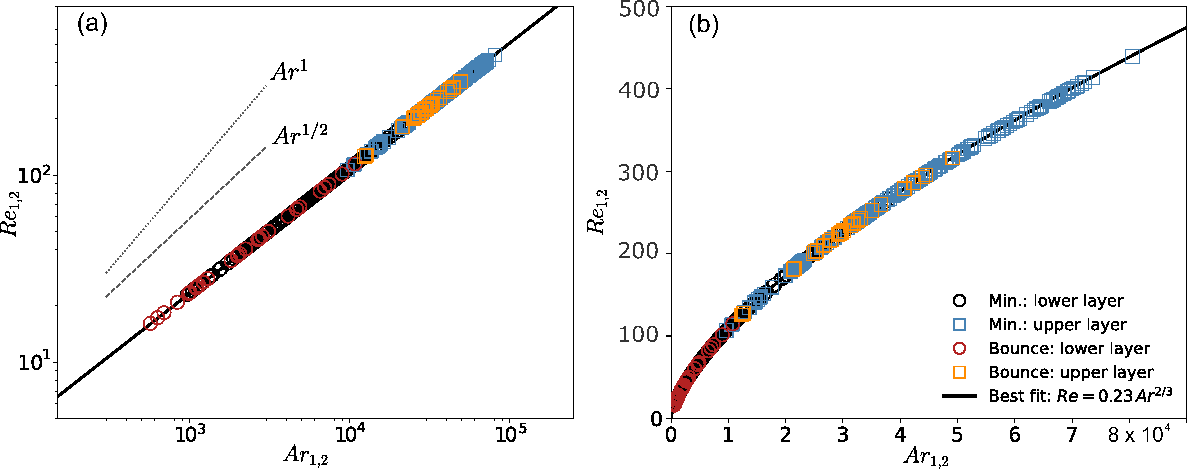}
	\caption{Drag law validation. Terminal-velocity Reynolds number $Re_{1,2}$ versus Archimedes number $Ar_{1,2}$ for all spheres in both fluid layers (log-log left, linear right). Solid line: best fit $Re = 0.23\,Ar^{2/3}$ ($R^2 > 0.99$). Triangles: Series~A (water--salt); circles: Series~B (water--glycerol); red: bouncing; blue: minimum.}
	\label{fig:drag_law}
\end{figure}

Terminal velocity in both fluid layers follows $Re = 0.23\,Ar^{2/3}$ (\autoref{fig:drag_law}), consistent with the intermediate-Reynolds-number scaling that underlies the drag law $C_d = \alpha\,Re^{-1/2}$ used in Eq.~\eqref{eq:drag_balance}. The fit holds across both Series~A and Series~B, confirming that a single universal prefactor~$\alpha$ is appropriate for both fluid systems. This cross-validation supports our results in comparison to the \cite{Srdic-Mitrovic1999, Magnaudet1997, Mandel2020}, among others. 

\subsection{Calibration of the drag prefactor \texorpdfstring{$\alpha$}{alpha}}
\label{subsec:alpha_calibration}

The drag prefactor~$\alpha$ in $C_d = \alpha\,Re^{-1/2}$ is not assumed \emph{a priori} but is calibrated against ground-truth data. Eight spheres with independently measured densities~$\rho_{s,\mathrm{meas}}$ are available. For each trial value of~$\alpha$, the recovery fit (Eq.~\ref{eq:exp_recovery}) is applied to all eight spheres and the inferred density~$\rho_{s,\mathrm{fit}}$ is computed from Eq.~\eqref{eq:rho_from_Uinf}. The optimal~$\alpha$ minimizes the root-mean-square error:
\begin{equation}\label{eq:alpha_opt}
	\alpha_\mathrm{opt} = \underset{\alpha \in [6,\,20]}{\arg\min}\;
	\sqrt{\frac{1}{N}\sum_{i=1}^{N}
		\bigl(\rho_{s,\mathrm{fit}}^{(i)} - \rho_{s,\mathrm{meas}}^{(i)}\bigr)^2}.
\end{equation}
The optimization yields an optimal value $\alpha_\mathrm{opt} = 11.68$, with a well-defined minimum: RMSE~$= 0.94$~kg/m$^3$, bias~$= +0.28$~kg/m$^3$, and maximum absolute error~$|\Delta\rho| = 1.87$~kg/m$^3$ over the eight spheres.

\subsection{Validation}

\begin{figure}[!ht]
	\centering
	\includegraphics[width=0.85\textwidth]{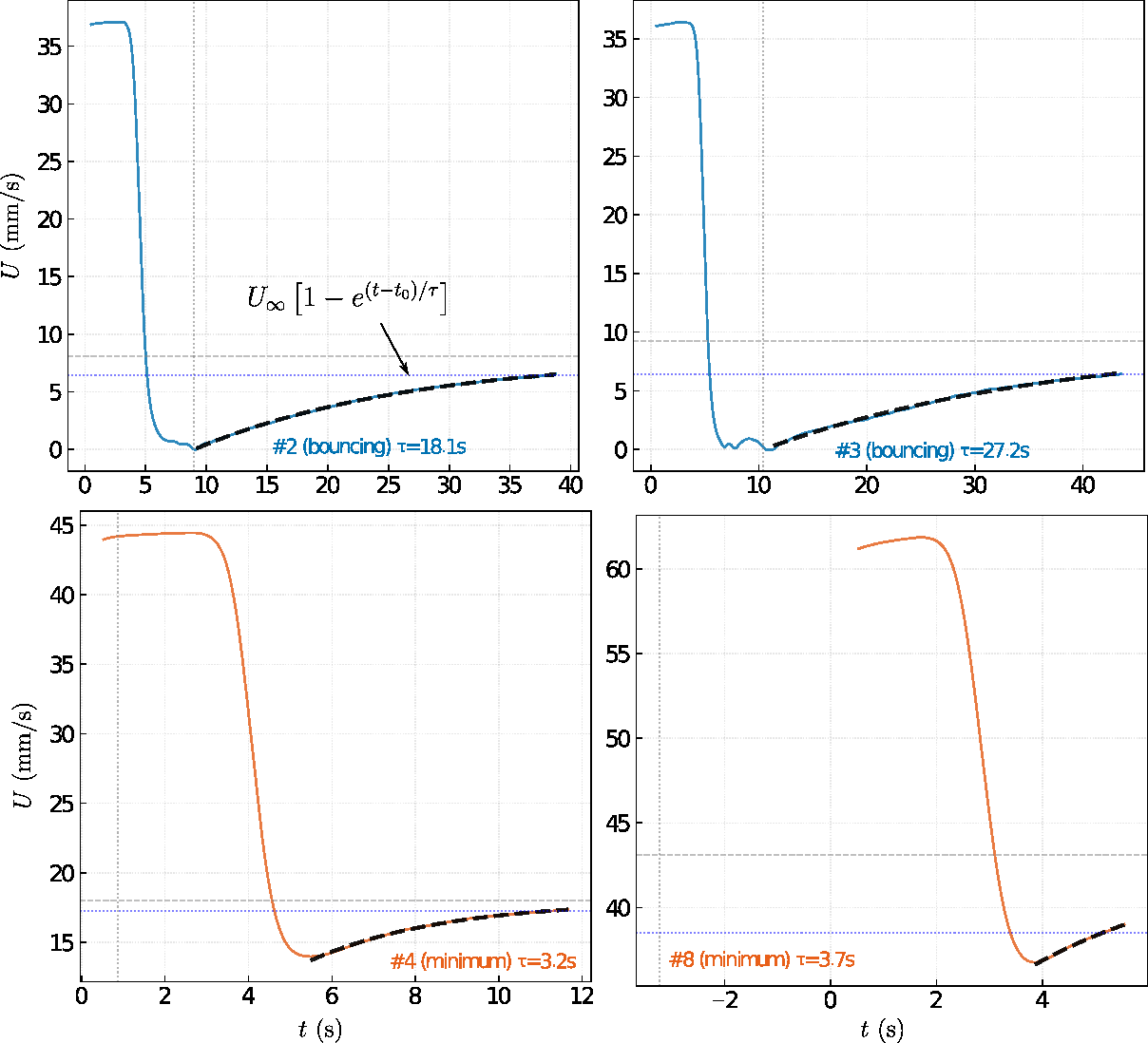}
	\caption{Exponential recovery fits (black dashed) to four out of eight spheres with measured densities. Each panel shows the measured speed (colored line), the fit $U(t) = U_\infty[1-\exp(-(t-t_0)/\tau)]$, and the virtual origin~$t_0$ (dotted vertical). Bouncing spheres (red) exhibit $\tau \approx 18$--$33$~s; minimum spheres (blue) have $\tau \approx 3$--$4$~s.}
	\label{fig:recovery_fits}
\end{figure}

\begin{figure}[!ht]
	\centering
	\includegraphics[width=\textwidth]{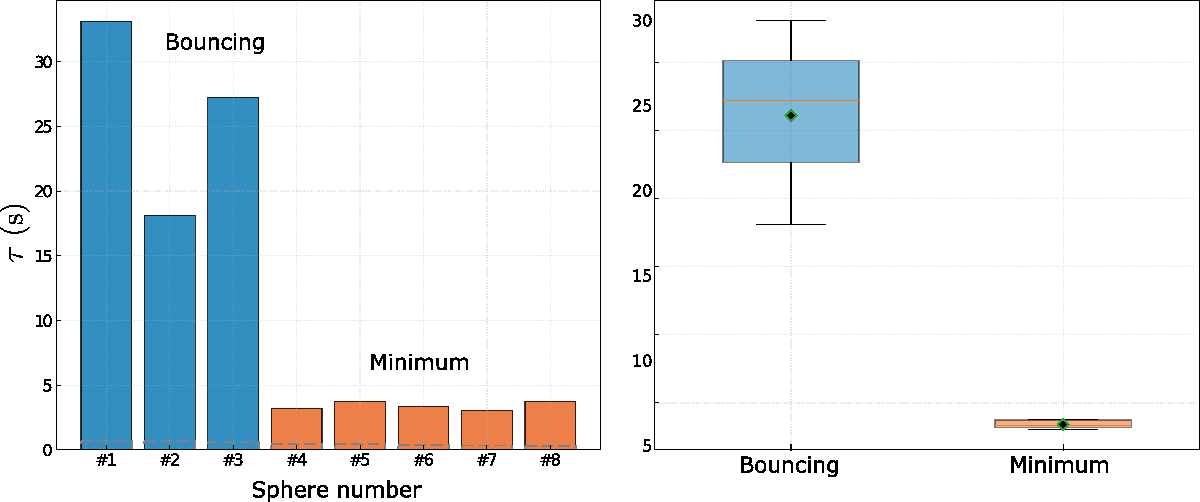}
	\caption{Recovery timescale~$\tau$ for the eight ground-truth spheres. (a)~Per-sphere bar chart: solid bars show the fitted~$\tau$, dashed outlines show the linearized drag timescale $\tau_\mathrm{drag}$. (b)~Box plots comparing the $\tau$ distributions of bouncing (red) and minimum (blue) spheres. The $\sim\!8\times$ separation reflects the different entrained film volumes.}
	\label{fig:tau_timescale}
\end{figure}

The calibrated model is validated in three ways (\autoref{fig:recovery_fits}):
\begin{enumerate}
	\item \textbf{Density recovery.} The fitted sphere densities reproduce the measured values with sub-1~kg/m$^3$ RMSE across all eight spheres.
	\item \textbf{Lower-layer terminal velocity.} The predicted $U_2 = U_\infty(1 - e^{-5})$ at $t = t_0 + 5\tau$ agrees with the measured lower-layer terminal velocity (mean ratio $U_{2,\mathrm{pred}}/U_{2,\mathrm{meas}} = 1.20$, reflecting partial recovery within the measurement window).
	\item \textbf{Upper-layer cross-check.} Using the same $\rho_{s,\mathrm{fit}}$ and $\alpha_\mathrm{opt}$, the predicted upper-layer terminal velocity reproduces the measured $U_1$ with mean ratio $U_{1,\mathrm{pred}}/U_{1,\mathrm{meas}} = 1.00$, confirming consistency of the drag law across both layers.
\end{enumerate}
The mean fit residual is $0.06$~mm/s, confirming that the single-parameter exponential model captures the recovery dynamics to within measurement uncertainty.

\end{document}